\documentclass[conference]{IEEEtran}
\IEEEoverridecommandlockouts
\usepackage{cite}
\usepackage{amsmath,amssymb,amsfonts}
\usepackage{algorithmic}
\usepackage{graphicx}

\usepackage{flushend}

\usepackage{url}

\usepackage{filecontents}

\usepackage{tikz}
\usetikzlibrary{positioning}
\usetikzlibrary{arrows,arrows.meta}
\usetikzlibrary{decorations.pathreplacing}
\usetikzlibrary{calc}

\newcommand\copyrighttext{%
  \footnotesize \textcopyright 2018 IEEE. Personal use of this material is permitted.
  Permission from IEEE must be obtained for all other uses, in any current or future
  media, including reprinting/republishing this material for advertising or promotional
  purposes, creating new collective works, for resale or redistribution to servers or
  lists, or reuse of any copyrighted component of this work in other works.
}
\newcommand\copyrightnotice{%
\begin{tikzpicture}[remember picture,overlay]
\node[anchor=south,yshift=10pt] at (current page.south) {\fbox{\parbox{\dimexpr\textwidth-\fboxsep-\fboxrule\relax}{\copyrighttext}}};
\end{tikzpicture}%
}

\usepackage{float}
\usepackage[nolist,nohyperlinks]{acronym}
\usepackage{textcomp}
\def\BibTeX{{\rm B\kern-.05em{\sc i\kern-.025em b}\kern-.08em
    T\kern-.1667em\lower.7ex\hbox{E}\kern-.125emX}}
\begin{document}

\title{CoRT: A Communication Robustness Testbed \\ for Industrial Control System Components}

\author{
\IEEEauthorblockN{Matthias Niedermaier}
\IEEEauthorblockA{\textit{Hochschule Augsburg} \\
Augsburg, Germany \\
Matthias.Niedermaier@hs-augsburg.de}
\and
\IEEEauthorblockN{Alexander von Bodisco}
\IEEEauthorblockA{\textit{Hochschule Augsburg} \\
Augsburg, Germany \\
Alexander.vonBodisco@hs-augsburg.de}
\and
\IEEEauthorblockN{Dominik Merli}
\IEEEauthorblockA{\textit{Hochschule Augsburg} \\
Augsburg, Germany \\
Dominik.Merli@hs-augsburg.de}
}

\maketitle
\copyrightnotice

\begin{abstract}
The number of interconnected devices is growing constantly due to rapid digitalization, thus providing attackers with a larger attack surface.
Particularly in critical infrastructures and manufacturing, where processes can be observed and controlled remotely, successful attacks could lead to high costs and damage.
Therefore, it is necessary to investigate \ac{ICS} devices like \acp{PLC} to make these sectors more secure.
One possible attack vector is the exploitation of the network communication of devices.
Thus, a robust communication system is essential to ensure security.
Unfortunately, the high demand for real-world \acp{ICS} makes it difficult to assess component security during its runtime. 
However, this is possible in a research testbed where tests could be done and analyzed in a safe environment.
In this paper, we introduce our testbed and measurement methods for communication robustness test research of \ac{ICS} components.

\end{abstract}


\acresetall
\section{Introduction}
Digitization is happening steadily, meaning that more and more devices are getting interconnected.
This trend of connected devices can be observed also in industrial environments and in critical infrastructure sectors.
Consequently, the attack surface over networks is also growing constantly.
A successful attack on industrial plants such as critical infrastructures results in high costs and damage.
Hence, the security of the devices used in these applications must be ensured and should be provided by the devices themselves.
Owing to remote access, security audits should also focus on robust and secure network communication.
However, a penetration test in a real-world production environment could lead to damage and outage.
Therefore, to analyze the communication robustness of these components, a testbed is necessary.

There are already a lot of testbeds for \ac{ICS} security research.
Holm et al. \cite{testbedsurvey} have analyzed 30 \ac{ICS} testbeds in 2015.
However, these mostly focus on analyzing a complete \ac{ICS} infrastructure, as opposed to testing single components.

In this paper, we present our testbed, focusing on the security of industrial components, especially the robustness of communication, 
which can influence the control behavior. 
The following requirements for the testbed, with corresponding measurement methods, have been defined:
\begin{itemize}
\item \textbf{Network capture}: The generated network traffic during attacks and tests must be captured for further investigation.
\item \textbf{Network reachability check}: It must be probed if the devices are reachable within the network during attacks.
\item \textbf{Electrical monitoring}: The electrical outputs must be monitored to recognize changes in the control behavior.
\item \textbf{Fast integration}: The integration of new devices into the testbed must be easy and fast.
\end{itemize}

The rest of the paper is structured as follows.
In Section \ref{sec:overview}, an overview of the testbed is given.
The devices currently deployed in the testbed are described in Section \ref{sec:devicesintestbed}.
Security tests, which can be done with CoRT, are explained in Section \ref{sec:experiments}. 
Finally, a conclusion is provided in Section \ref{sec:conclusion}.

\section{CoRT Testbed} \label{sec:overview}
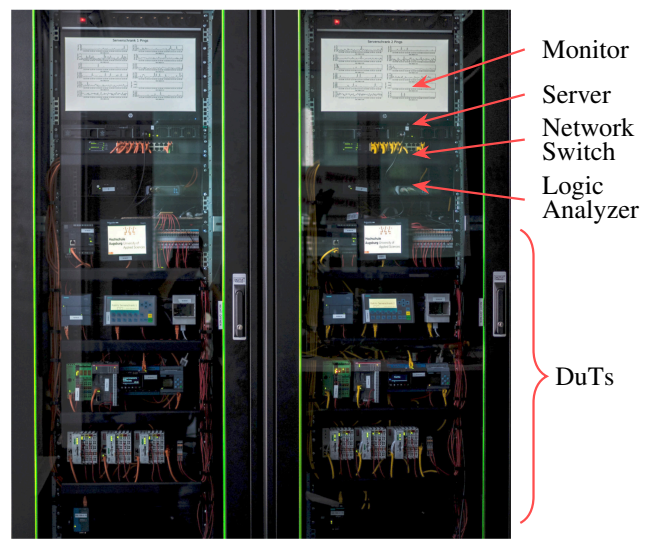
\begin{figure}[H]
\center
\begin{tikzpicture}
	\node[inner sep=0pt] (russell) at (0,0)
    {\includegraphics[width=.75\columnwidth]{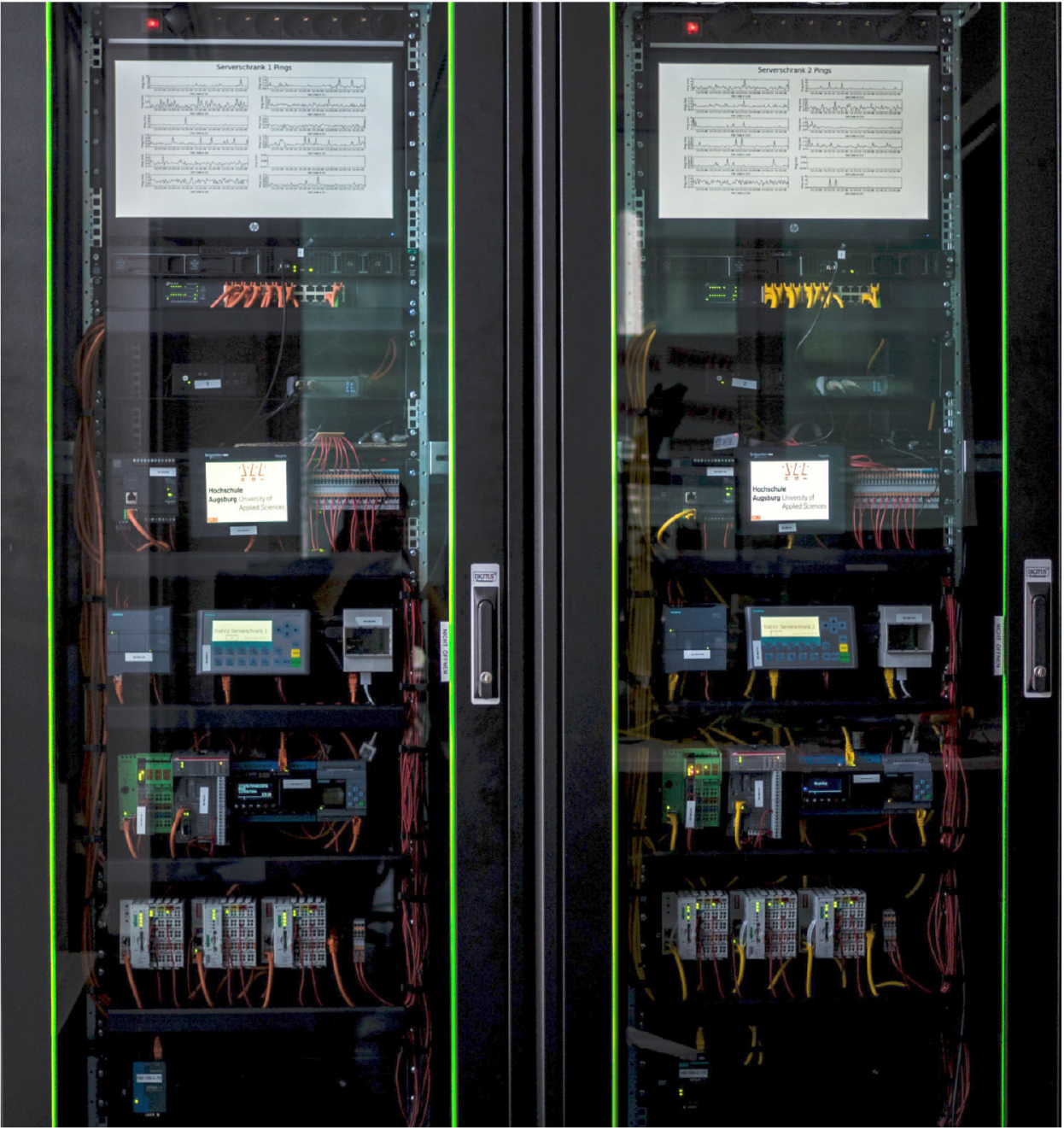}};
    
    \draw [{Stealth[scale=1.0]}-, color=red!70!white, line width=0.3mm]  (2,2.5) to (3.5,3) node[right=0.1cm, text width=1.0cm]{\textcolor{black}{Monitor}};
    \draw [{Stealth[scale=1.0]}-, color=red!70!white, line width=0.3mm]  (2,2.0) to (3.5,2.4) node[right=0.1cm, text width=1.0cm]{\textcolor{black}{Server}};
    \draw [{Stealth[scale=1.0]}-, color=red!70!white, line width=0.3mm]  (2,1.6) to (3.5,1.8) node[right=0.1cm, text width=1.0cm]{\textcolor{black}{Network \\[-0.1cm] Switch}};
    \draw [{Stealth[scale=1.0]}-, color=red!70!white, line width=0.3mm]  (2,1.2) to (3.5,1.0) node[right=0.1cm, text width=1.0cm]{\textcolor{black}{Logic \\[-0.1cm] Analyzer}};
    
    \draw [decorate,decoration={brace,amplitude=10pt,raise=4pt},yshift=0pt,color=red!70!white, line width=0.3mm] (3.3,0.6) -- (3.3,-3.3) node [black,midway,xshift=1.0cm] {\textcolor{black}{DuTs}};
\end{tikzpicture}
\caption{Picture of the Testbed Built into Racks}
\label{fig:rackpic}
\end{figure}

A communication robustness test measures the steadiness of control signals under various communication parameters and loads.
Fig. \ref{fig:rackpic} shows our testbed divided into two identical racks, enabling comparison of the results.

At the bottom are the \acp{DuT}, such as \acp{PLC}, \acp{HMI}, and bus couplers.
The electrical outputs of these are wired to a logic analyzer.
All \acp{DuT} are interconnected with a network switch to a server.
Furthermore, there is a monitor built in to keep tabs on the analyzed data.
The \acp{DuT} are mounted on an EN 50022 rail, which is commonly used in the industrial sector.
Both racks are lockable and have wheels for easy transportation.

\subsection{Attacker Model}
This testbed is focused on, but not limited to, two attacker models:
a) an attacker who has remote access to the network and 
b) an attacker who has local access to the \ac{ICS} components with basic knowledge.
Both of them are able to inject network traffic, e.g. to send commands or perform a \ac{MitM} attack.
Attacker model b is furthermore able to locally manipulate input signals and has direct network access to the \ac{DuT}.

\subsection{Robustness of Devices}
The robustness of an \ac{IIoT} device is essential, because they mostly control machines and interact with their environment.
Therefore, an outage or loss of control creates a problem.
With the CoRT, the research question of how network communication could influence the robustness of industrial systems can be analyzed.
The communication robustness of different industrial components can be measured with fuzz testing frameworks.
Besides, these must be specialized for proprietary protocol fuzzing in \acp{ICS}, such as PropFuzz \cite{niedermaier2017propfuzz}.

\subsection{Schematic Overview of the Testbed}
\begin{figure}[H]
\centering
\begin{tikzpicture}[node distance=1cm,
    auto,
    block/.style={
      rectangle,
      draw=black,
      align=center,
      rounded corners,
      dashed
    }
  ]
  \coordinate (a) at (1,1.8);
  \coordinate (b) at (2.8,1.8);
  \coordinate (c) at (4.6,1.8);
  \coordinate (d) at ([xshift=2.4cm, yshift=0.0cm]b.east);
  \coordinate (e) at ([xshift=1.8cm, yshift=0.0cm]d.east);
  
  \draw [-, line width=0.5mm, color=red] ([xshift=0.0cm, yshift=0.0cm]a) -- ([xshift=2.2cm, yshift=0.0cm]c);
  \node[rectangle, align=center, minimum width=1.5cm, minimum height=0.6cm, anchor=west] at ([xshift=1.2cm, yshift=0.25cm]c)(g) {\textcolor{red}{\small 24V}};

  \node[block, draw, align=center, minimum width=1.5cm, minimum height=0.6cm, anchor=west, fill=white] at (a.north) (x) {\acs{DuT}\_0};
  \node[block, draw, align=center, minimum width=1.5cm, minimum height=0.6cm, anchor=west, fill=white] at (b.north) (y) {\acs{DuT}\_1};
  \node[block, draw, align=center, minimum width=1.5cm, minimum height=0.6cm, anchor=west, fill=white] at (c.north) (z) {\acs{DuT}\_N};
  
  \draw [-, line width=0.5mm, dashed] ([xshift=-1.1cm, yshift=-0.6cm]x.south) -- ([xshift=0.8cm, yshift=-0.6cm]z.south);  
  \node[rectangle, align=center, minimum width=1.5cm, minimum height=0.6cm, anchor=south west] at ([xshift=-1.9cm, yshift=-0.7cm]z.south)(h) {\small Network};
  \draw [-, line width=0.5mm, dashed] ([xshift=-0.2cm, yshift=0.0cm]x.south) -- ([xshift=-0.2cm, yshift=-0.6cm]x.south);
  \draw [-, line width=0.5mm, dashed] ([xshift=-0.2cm, yshift=0.0cm]y.south) -- ([xshift=-0.2cm, yshift=-0.6cm]y.south);
  \draw [-, line width=0.5mm, dashed] ([xshift=-0.2cm, yshift=0.0cm]z.south) -- ([xshift=-0.2cm, yshift=-0.6cm]z.south);

  \node[block, draw, align=center, minimum width=0.5cm, minimum height=3.2cm, anchor=west, text width=0.4cm] at ([xshift=-1.0cm, yshift=0.4cm]a.north) (u) {S \\ e \\ r \\ v \\ e \\ r};
  
  \node[rectangle, draw, align=center, minimum width=2.5cm, minimum height=0.6cm, anchor=south, text width=2.4cm] at ([xshift=0.0cm, yshift=0.8cm]y.north) (t) {Logic Analyzer};
  \draw [-, line width=0.5mm, dotted, color=blue] ([xshift=0.0cm, yshift=0.0cm]x.north) -- ([xshift=-0.6cm, yshift=0.0cm]t.south);
  \draw [-, line width=0.5mm, dotted, color=blue] ([xshift=0.0cm, yshift=0.0cm]y.north) -- ([xshift=0.0cm, yshift=0.0cm]t.south);
  \draw [-, line width=0.5mm, dotted, color=blue] ([xshift=0.0cm, yshift=0.0cm]z.north) -- ([xshift=0.6cm, yshift=0.0cm]t.south);
  \node[rectangle, align=center, minimum width=1.5cm, minimum height=0.6cm, anchor=west] at ([xshift=0.65cm, yshift=-0.2cm]t.south)(e) {\textcolor{blue}{\small Wires}};
  
  \draw [-, line width=0.25mm] (t.west) -| ([xshift=0.0cm, yshift=1.0cm]u.east);
 
  \draw [decorate,decoration={brace,amplitude=5pt,raise=4pt},yshift=0pt,color=red!70!white, line width=0.3mm] (6.9,3.5) -- (6.9, 2.1) node [black,midway,xshift=0.25cm, text width=1.2cm] {\textcolor{black}{Critical \\ Control \\ Part}};
  \draw [decorate,decoration={brace,amplitude=5pt,raise=4pt},yshift=0pt,color=red!70!white, line width=0.3mm] (6.9,1.7) -- (6.9, 0.5) node [black,midway,xshift=0.25cm, text width=1.2cm] {\textcolor{black}{Commu-nication \\ Part}};
   
  \end{tikzpicture}
  \caption{Schematic Overview of the Testbed}
  \label{fig:setup}
\end{figure}
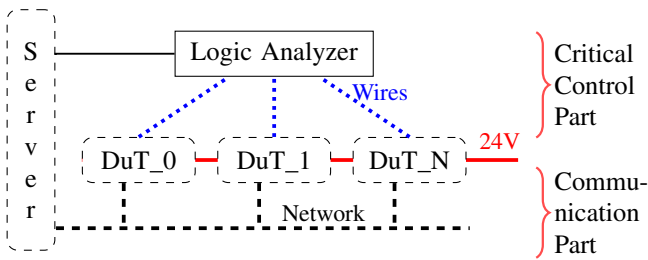

Fig. \ref{fig:setup} gives a schematic overview of the testbed components.
On the server, there are three \acp{VM} for measuring, programming, and attacking the \acp{DuT}.
In order not to influence the measurement, this separation is necessary, because some attack tools produce a high system load.
Furthermore, additional attack tools and further \acp{DuT} could be integrated into the testbed.

The \acp{DuT} are powered by 24V, which can be switched on and off remotely by the measurement \ac{VM}.
With the logic analyzer, the critical control part is observed.

\subsection{Electrical Behavior}
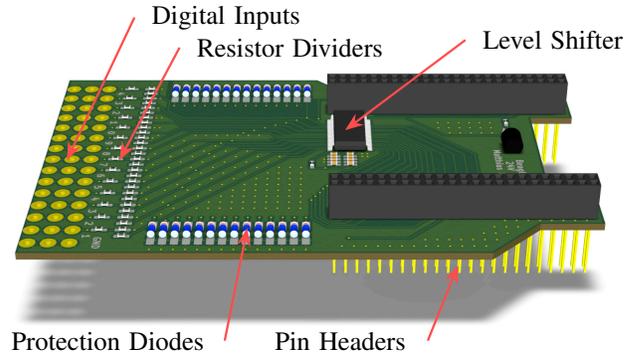
\begin{figure}[H]
\center
\begin{tikzpicture}
\node[inner sep=0pt] (russell) at (0,0)
    {\includegraphics[width=.95\columnwidth]{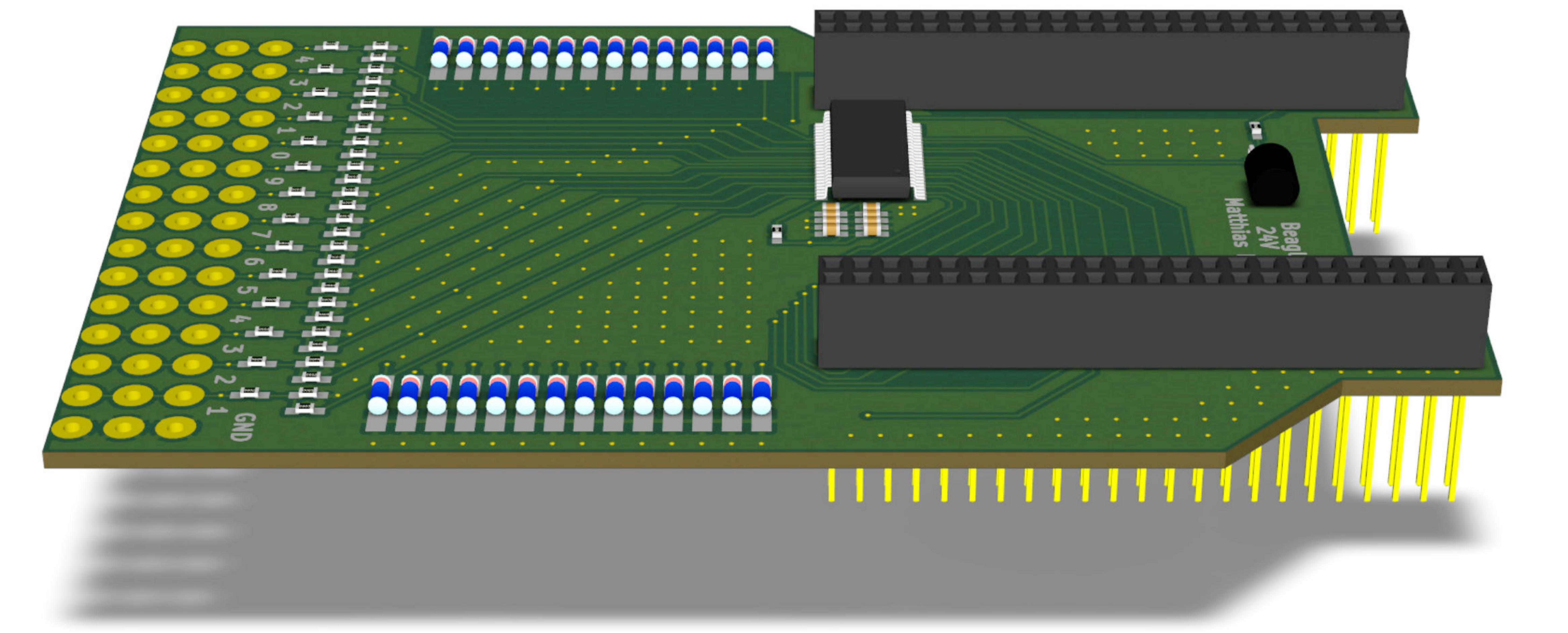}};
    \draw [{Stealth[scale=1.0]}-, color=red!70!white, line width=0.3mm]  (-3.3,0.5) to (-2.4,2.4) node[right=0.1cm, text width=2.0cm]{\textcolor{black}{Digital Inputs}};
    \draw [{Stealth[scale=1.0]}-, color=red!70!white, line width=0.3mm]  (-2.6,0.5) to (-1.8,2.0) node[right=0.1cm, text width=2.8cm]{\textcolor{black}{Resistor Dividers}};
    \draw [{Stealth[scale=1.0]}-, color=red!70!white, line width=0.3mm]  (0.4,0.9) to (2.0,2.1) node[right=0.1cm, text width=2.0cm]{\textcolor{black}{Level Shifter}};
    \draw [{Stealth[scale=1.0]}-, color=red!70!white, line width=0.3mm]  (-0.9,-0.4) to (-1.2,-1.9) node[left=-0.1cm, text width=2.8cm]{\textcolor{black}{Protection Diodes}};
    \draw [{Stealth[scale=1.0]}-, color=red!70!white, line width=0.3mm]  (1.9,-0.9) to (1.5,-1.9) node[left=-0.1cm, text width=2.0cm]{\textcolor{black}{Pin Headers}};
\end{tikzpicture}
\vspace*{-0.2cm}
\caption{Rendered Image of the PCB of the Logic Adapter Board}
\label{fig:beaglelogic}
\end{figure}

To measure the robustness of the \acp{DuT}, the electrical outputs are observed with a logic analyzer. 
All \acp{PLC} are configured to toggle every single cycle, resulting in a frequency between 20Hz and 20kHz depending on the device.
These signals are measured with a logic analyzer and logged on the measurement \ac{VM}.
For the logic analyzer task, a BeagleBone Green running Beaglelogic \cite{beaglelogic} with a custom \ac{PCB} is used.
Using BeagleLogic, it is possible to measure a sample frequency up to 100MHz.
Fig. \ref{fig:beaglelogic} shows the adapter board designed by us.

This is necessary to measure up to 30V and convert it to the 3.3V input signal of the BeagleBone Green.
On the \ac{PCB}, there are mainly resistor dividers, protection diodes, and an SN74LVCH16245A \cite{texas} 16-Bit level shifter.
The left side of the \ac{PCB} is connected to the outputs of the \acp{DuT} and the pin header is mounted on the BeagleBone Green.
With the logic analyzer setup, influences on the control behavior of \acp{PLC} are measured and logged.

\subsection{Ping Response}
To ensure network reachability during the tests, the ping response of all devices is measured. 
This is done with a fping \cite{fping} script every 100ms, which has been tested as a trade-off between measuring accuracy and not influencing the devices by flooding.
If a network stack on a \ac{DuT} crashes, it is detected and logged.

\subsection{Network Capture}
To analyze tests in detail, a network capture of this period is essential. The complete network traffic within the testbed is mirrored on one port and captured with tcpdump \cite{jacobson1989tcpdump} in a log rotation. 
For example, if a device fails during an attack, the timestamped and corresponding captures are used for further investigation.

\subsection{Visualization}
\begin{figure}[H]
\center
\vspace*{-0.3cm}
\begin{tikzpicture}
\node[inner sep=0pt] (russell) at (0,0)
    {\includegraphics[width=.95\columnwidth]{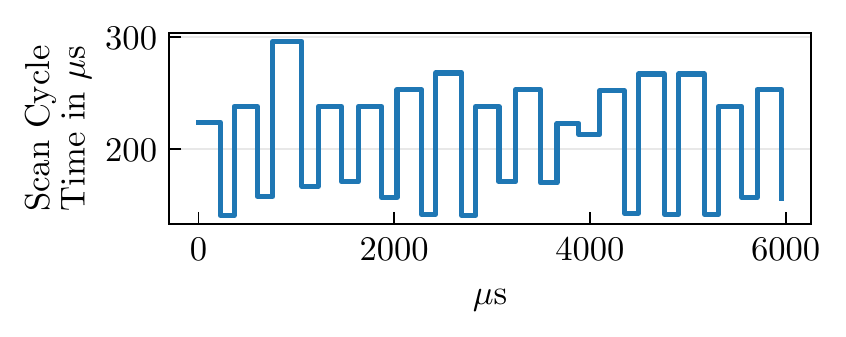}};
\end{tikzpicture}
\vspace*{-0.3cm}
\caption{Visualization of the Scan Cycle Time of a Siemens S7-1211C}
\label{fig:plot}
\end{figure}

Besides the logging, the measurement results are directly visualized on monitors on the testbed. 
The cycle time \cite{mader2000classification} of a Siemens S7-1211C during idle is illustrated in Fig. \ref{fig:plot}. 
But this is not constant, because it is influenced by communication and housekeeping during the execution.
One scan cycle takes from 140$\mu$s to 300$\mu$s during idle measurement.
This can later be compared with the scan cycle time during an attack, e.g. to check if the \ac{DuT} is being influenced. 

Furthermore, the ping response, current test results, and statistics of the network capture are visualized. This enables a quick overview at any time.

\subsection{Virtualized Devices}
For the measurement of the response time of a command sent over the network to the \ac{DuT}, it is necessary to virtualize components, e.g. sensors or other \ac{IIoT} devices.
The control command (e.g. set an output over an industrial protocol) is sent by the measurement \ac{VM}, which reduces timing dependencies during the measurement.
Thereafter, the electrical output is measured by the logic analyzer and is processed.
In order to send a command, the industrial protocol, e.g. BACnet and Modbus, must be used and implemented into the test.

\subsection{Input Signal Generation}
To simulate electric input signals for the \acp{DuT}, these must be generated to be read by real hardware.
This is realized with a \ac{USB}-to-serial converter.
Both the generated input and output signals of the \ac{DuT} are measured by the logic analyzer.
The delay between them is the response time.
A jitter of this could be caused if, for example, the \ac{DuT} faces a high \ac{CPU} load due to network communication.

\section{\acp{DuT} in the Testbed} \label{sec:devicesintestbed}
For the CoRT testbed, diverse vendors and products have been chosen.
With this, it is possible to compare implementations based on their security level from a technical point of view. 
However, single devices can also be tested.
Table \ref{tab_racksetup} lists currently employed devices with a selection of open ports.

\begin{table*}[!htbp]
\centering
\caption{\acp{PLC} Currently Employed within the Testbed}
\label{tab_racksetup}
\resizebox{\textwidth}{!}{%
\begin{tabular}{l l l | l | l | l}
\hline \hline
\textbf{Vendor}           & \textbf{Product}        & \textbf{Vendor No.} & \textbf{IP Rack 1}   & \textbf{IP Rack 2} & \textbf{Selection of Open Ports} \\           
\hline  
Siemens                   & CPU 1211C               & 6ES7211-1AE40-0XB0  & 192.168.0.10         & 192.168.0.110      & 80, 102, 443 \\ 
Siemens                   & KP 300                  & 6AV6647-0AH11-3AX0  & 192.168.0.11         & 192.168.0.111      & 102, 2308 \\
Phoenix                   & ILC 151                 & 2700974             & 192.168.0.20         & 192.168.0.120      & 21, 80, 1962, 41100 \\
ABB                       & PM554-T                 & 1SAP120600R0071     & 192.168.0.21         & 192.168.0.121      & 21, 502, 1200, 1201 \\
Crouzet                   & em4 B26-2GS             & 88981133            & 192.168.0.22         & 192.168.0.122      & 502, 42424 \\
Siemens                   & LOGO! 24RCE             & 6ED1052-1CC01-0BA8  & 192.168.0.23         & 192.168.0.123      & 80, 102, 502, 8080 \\
Wago                      & Controller KNX IP       & 750-889             & 192.168.0.30         & 192.168.0.130      & 21, 80, 443, 502, 2455, 6626 \\
Wago                      & Controller PFC100       & 750-8100            & 192.168.0.31         & 192.168.0.131      & 22, 80, 443, 502, 4840, 6626, \\
                          &                         &                     &                      &                    & 11740 \\
Wago                      & Controller ETHERNET     & 750-880             & 192.168.0.32         & 192.168.0.132      & 21, 80, 443, 502, 2455, 6626, \\
                          &                         &                     &                      &                    & 44818 \\
Wago                      & Controller BACnet/IP    & 750-831             & 192.168.0.33         & 192.168.0.133      & 21, 80, 443, 502, 2455, 6626,  \\
                          &                         &                     &                      &                    & 47808 \\
Schneider                 & TM221CE16T              & TM221CE16T          & 192.168.0.50         & 192.168.0.150      & 502, 44818 \\
Schneider                 & HMISTU855               & HMISTU855           & 192.168.0.51         & 192.168.0.151      & 502, 6001 \\
OpenPLC v2 \cite{alves2014openplc} & Raspberry Pi 3 & Commit f1a2645      & 192.168.0.60         & 192.168.0.160      & 22, 502, 8080, 20000 \\
Moxa                      & NP5110                  & NP5110              & 192.168.0.70         & 192.168.0.170      & 23, 80, 443, 950, 966, 4900 \\
\end{tabular}
} 
\vspace*{-0.3cm}
\end{table*}

\subsection{Common Network Protocols}
Industrial components often use common network protocols for tasks such as monitoring and visualization.
On the testbed, we have the following common network protocols:
\begin{itemize}
\item \textbf{\ac{FTP} "Port: 21"} is used for file transfer.
Within industrial networks, this is used for logging and updating the firmware, for example.
\item \textbf{\ac{SSH} "22"} refers to a network protocol that can be used to securely communicate with a remote device. 
It is often used to make a remote command line available locally if, for example, the \ac{PLC} runs Linux.
\item \textbf{Telnet "23"} is a client/server protocol that is based on a character-oriented data exchange over a TCP connection.
\acp{PLC} could be partly configured over it.
\item \textbf{\ac{HTTP} "80"} is mainly used to load web pages (hypertext files) into a web browser.
\textbf{\ac{HTTPS} "443"} uses an additional transport security.
These are used for the visualization of diagnostic information and sensor values.
\end{itemize}

\subsection{Industrial Network Protocols}
In modern plants, fieldbuses are increasingly being replaced by an IP-based communication system.
Within CoRT, five common industrial protocols on real hardware are available.

\begin{itemize}
\item The \textbf{Modbus/TCP "502"} protocol is a communication protocol based on a master/slave architecture.
\item \textbf{KNX IP "3671"} is mostly used for building automation.
\item \textbf{OPC Unified Architecture "4840"} is an industrial \ac{M2M} communication protocol that works across manufacturers.
\item \textbf{Ethernet/IP "44818"} is a real-time Ethernet mainly used in automation technology.
\item \textbf{BACnet/IP "47808"} is mostly used in building automation, ensuring interoperability between devices of different manufacturers, if all partners agree on certain building blocks defined by the standard.
\end{itemize}

\subsection{Proprietary Network Protocols}
Furthermore, there are proprietary protocols, which are only partly understood.
The \textbf{S7comm "102"} protocol is used by Siemens devices to communicate, for example, with the \ac{IDE} and \acp{HMI}.
Moreover, other vendors use proprietary protocols to program \acp{PLC} and \acp{HMI} (\mbox{\textbf{Phoenix Contacts "1962, 41100"}},
\mbox{\textbf{ABB "1200, 1201"},}
\mbox{\textbf{WinCC "2308"},}
\mbox{\textbf{WAGO-Service-Protocol "2455, 6626"},}
\mbox{\textbf{Crouzet "42424"},}
\mbox{\textbf{Codesys "11740"}).}
In terms of security, these protocols are particularly interesting, because they execute privileged commands, such as setting the run mode of a device and updating the user application.

\section{Experiments with CoRT} \label{sec:experiments}
With CoRT, a playground for researchers, organizations, and academic collaborators is made available.
It is nearly impossible to make tests in an operating \ac{ICS}, let alone change hardware or software.
Even if tests can be performed, the bulk of the necessary data is not recorded and cannot be evaluated.

\begin{figure}[H]
\centering
\vspace*{-0.3cm}
\begin{tikzpicture}[node distance=1cm,
    auto,
    block/.style={
      rectangle,
      draw=black,
      align=center,
      rounded corners,
      dashed
    }
  ]
  \coordinate (a) at (1,5.0);
  \coordinate (b) at (1,4.2);
  \coordinate (c) at (1,3.4);
  \coordinate (d) at (1,2.6);
  \coordinate (e) at (1,1.8);
  \coordinate (f) at (1,1);
  
  \node[block, draw, align=center, minimum width=3.6cm, minimum height=0.5cm, anchor=west] at (a.north) (q) {Start};  
  \node[block, draw, align=center, minimum width=3.6cm, minimum height=0.5cm, anchor=west] at (b.north) (r) {Power Cycle};
  \node[block, draw, align=center, minimum width=3.6cm, minimum height=0.5cm, anchor=west] at (c.north) (s) {Begin Measurement};
  \node[rectangle, align=center, minimum width=1.5cm, minimum height=0.5cm, anchor=west] at ([xshift=0.3cm, yshift=-0.3cm]s.east)(x) {\small (1) Idle Measurement};
  \node[block, draw, align=center, minimum width=3.6cm, minimum height=0.5cm, anchor=west] at (d.north) (u) {Test/Attack};
  \node[rectangle, align=center, minimum width=1.5cm, minimum height=0.5cm, anchor=west] at ([xshift=0.3cm, yshift=-0.0cm]u.east)(y) {\small (2) Attack Measurement};
  \node[rectangle, align=center, minimum width=1.5cm, minimum height=0.5cm, anchor=west] at ([xshift=0.3cm, yshift=-0.5cm]u.east)(z) {\small (3) Post Measurement};
  \node[block, draw, align=center, minimum width=3.6cm, minimum height=0.5cm, anchor=west] at (e.north) (v) {End Measurement};
  \node[block, draw, align=center, minimum width=3.6cm, minimum height=0.5cm, anchor=west] at (f.north) (w) {Analyze};

  \draw [-{Stealth[scale=1.0]}]  (q.south) to (r.north);
  \draw [-{Stealth[scale=1.0]}]  (r.south) to (s.north);
  \draw [-{Stealth[scale=1.0]}]  (s.south) to (u.north);
  \draw [-{Stealth[scale=1.0]}]  (u.south) to (v.north);
  \draw [-{Stealth[scale=1.0]}]  (v.south) to (w.north);
  
  \draw [->,color=black!50!white] (s.east) to [out=0,in=90] (x.north);
  \draw [->,color=black!50!white] (z.south) to [out=270,in=0] (v.east);

  \draw [->,color=black!50!white] (w.south) to [out=-120,in=-90] ($(w)+(-2.2,0)$) to [out=90,in=-90] ($(q)+(-2.2,0)$) to [out=90, in=130 ] (q.north);
  \node[rectangle, align=center, minimum width=1.5cm, minimum height=0.6cm, anchor=west] at ([xshift=-0.2cm, yshift=-0.3cm]w.south)(v) {\small Next Test or End};
  \end{tikzpicture}
  \vspace*{-0.5cm}
\caption{Illustration of a Test Sequence}
\label{fig:testsequence}
\end{figure}
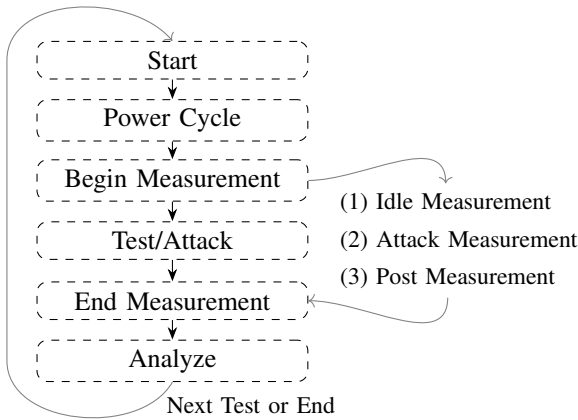

To evaluate different \acp{DuT}, predefined sequences of tests are used.
One of these test sequences is illustrated in Fig. \ref{fig:testsequence}.
After it starts, an automated power cycle of the \ac{DuT} can be done.
This becomes necessary if previous tests have influenced the \ac{DuT} or if it does not recover, e.g. after a successful \ac{DoS} attack.
Afterward, the measurement begins, including network captures, continuous reachability check, and electrical output measurement.
The measurement process has three phases: (1) pre-idle, (2) attack, and (3) post-idle monitoring.

During the test/attack, the \ac{DuT} is penetrated with different kinds of network robustness tests and vulnerability scans. 
Finally, the results are analyzed, logged, and are visualized on the monitor.
If additional tests are in the queue, the sequence starts from the beginning until all tests are finished.
It is possible to perform a full test on a single device, or to apply one test on every device in the rack. 
This depends on the type of validation, such as fuzzing of a single protocol on a device, which are not supported by other \acp{DuT}. 
Our first measurements indicate that network traffic can influence devices.



\section{Conclusion} \label{sec:conclusion}
The proposed testbed is a combination of network devices, measurement equipment, and industrial components. It allows studying \ac{ICS} devices like \acp{PLC}, sensors, and \acp{HMI} in detail without influencing real-world processes.
Furthermore, the testbed measurements are automatically recorded and analyzed, where new test scenarios can build up, without having to worry about the measurement setup.
With this testbed, a basis for communication robustness tests has been built.


\section*{Acknowledgment}
The work of the CoRT forms part of the RiskViz \cite{RiskViz} research project, which is funded by the \ac{BMBF}, with the aim of creating \ac{ICS} risk map.


\bibliographystyle{ieeetr}
\bibliography{\jobname}

\begin{acronym}
 \acro{A}{Availability}
 \acro{ACK}{acknowledgment}
 \acro{ARP}{Address Resolution Protocol}
 \acro{BMBF}{Federal Ministry of Education and Research}
 \acro{C}{Confidentiality}
 \acro{CIA}{Confidentiality, Integrity and Availability}
 \acro{CID}{Company IDentifier}
 \acro{CPE}{Common Platform Enumeration}
 \acro{CPU}{Central Processing Unit}
 \acro{CRT}{Communication Robustness Test}
 \acro{CSV}{Comma-Separated Values}
 \acro{CVE}{Common Vulnerabilities and Exposures}
 \acro{DCS}{Distributed Control System}
 \acrodefplural{DCS}{Distributed Control Systems}
 \acro{DHCP}{Dynamic Host Configuration Protocol}
 \acro{DoS}{Denial of Service}
 \acro{DuT}{Device under Test}
 \acrodefplural{DuT}{Devices under Test}
 \acro{ERP}{Enterprise Resource Planning}
 \acro{FTP}{File Transfer Protocol}
 \acro{HTTP}{Hypertext Transfer Protocol}
 \acro{HTTPS}{Hypertext Transfer Protocol Secure}
 \acro{HMI}{Human Machine Interface}
 \acrodefplural{HMI}{Human Machine Interfaces}
 \acro{I}{Integrity}
 \acro{IEEE}{Institute of Electrical and Electronics Engineers}
 \acro{ICS}{Industrial Control System}
 \acrodefplural{ICS}{Industrial Control Systems}
 \acro{IDE}{Integrated Development Environment}
 \acro{IoT}{Internet of Things}
 \acro{IIoT}{Industrial Internet of Things}
 \acro{IP}{Internet Protocol}
 \acro{M2M}{Machine to Machine}
 \acro{MAC}{Media Access Control}
 \acro{MES}{Manufacturing Execution System}
 \acro{MitM}{Man-in-the-Middle}
 \acro{NSE}{Nmap Scripting Engine}
 \acro{OS}{Operating System}
 \acro{OSI}{Open Systems Interconnection}
 \acro{OT}{Operational Technology}
 \acro{OUI}{Organizationally Unique Identifier}
 \acro{pcap}{packet capture}
 \acrodefplural{pcap}{packet captures}
 \acro{PCB}{Printed Circuit Board}
 \acro{PLC}{Programmable Logic Controller}
 \acrodefplural{PLC}{Programmable Logic Controllers}
 \acro{PoC}{Proof of Concept}
 \acro{RA}{Registration Authority}
 \acro{SCADA}{Supervisory Control and Data Acquisition}
 \acro{SSH}{Secure Shell}
 \acro{SYN}{synchronize}
 \acro{TAP}{Terminal Access Point}
 \acro{TCP}{Transmission Control Protocol}
 \acro{ToS}{Type of Service}
 \acro{TTL}{Time to Live}
 \acro{USB}{Universal Serial Bus}
 \acro{VLAN}{Virtual Local Area Network}
 \acrodefplural{VLAN}{Virtual Local Area Networks}
 \acro{VM}{Virtual Machine}
 \acro{VPN}{Virtual Private Network}
\end{acronym}

\end{document}